\documentclass[runningheads]{llncs}
\usepackage{authblk}
\usepackage{url}
\usepackage{hyperref}
\usepackage{algorithm}
\usepackage{graphicx}
\usepackage{threeparttable}
\usepackage{booktabs}
\usepackage[noend]{algpseudocode}
\usepackage[T1]{fontenc}
\usepackage{graphicx}
\begin{document}
\title{Deep Neural Network Pruning for Nuclei Instance Segmentation in Hematoxylin \& Eosin-Stained Histological Images}
\titlerunning{DNN Pruning for Nuclei Instance Segmentation}
%


\author{Amirreza Mahbod\thanks{The first two authors contributed equally to this work.}\inst{1} \and
	Rahim Entezari\inst{2,3} \and
	Isabella Ellinger\inst{1} \and
	Olga Saukh\inst{2,3}
}
\authorrunning{A. Mahbod and R. Entezari et al.}

\institute{ Institute for Pathophysiology and Allergy Research, Medical University of Vienna, Vienna, Austria \and
	Institute of Technical Informatics, Technical University Graz, Graz, Austria
	\\
	\and
	 Complexity Science Hub, Vienna, Austria\\
}

\maketitle              
\begin{abstract}
Recently, pruning deep neural networks (DNNs) has received a lot of attention for improving accuracy and generalization power, reducing network size, and increasing inference speed on specialized hardwares. Although pruning was mainly tested on computer vision tasks, its application in the context of medical image analysis has hardly been explored. This work investigates the impact of well-known pruning techniques, namely layer-wise and network-wide magnitude pruning, on the nuclei instance segmentation performance in histological images. Our utilised instance segmentation model consists of two main branches: (1) a semantic segmentation branch, and (2) a deep regression branch. We investigate the impact of weight pruning on the performance of both branches separately, and on the final nuclei instance segmentation result. Evaluated on two publicly available datasets, our results show that layer-wise pruning delivers slightly better performance than network-wide pruning for small compression ratios (CRs) while for large CRs, network-wide pruning yields superior performance. For semantic segmentation, deep regression and final instance segmentation, 93.75\,\%, 95\,\%, and 80\,\% of the model weights can be pruned by layer-wise pruning with less than 2\,\% reduction in the performance of respective models.

\end{abstract}

\section{Introduction}
\label{sec:intro}

Deep learning-based approaches have shown excellent performances for various computer vision tasks such as classification, detection, and segmentation problems~\cite{khan2018guide}. They have also been extensively used in various medical image analysis settings~\cite{anwar2018medical}. While deep neural network (DNN) algorithms perform better than other image processing or classical machine learning approaches,  they comprise millions of trainable parameters that slow down training and inference, and need powerful computational and storage resources~\cite{9111831}.

In recent years, various pruning techniques have been proposed, and their impact on DNN performance was explored. As shown in former studies, pruning can increase generalization capability, DNN performance, and at times decrease inference time~\cite{hoefler2021sparsity}.  

Pruning methods can be generally classified into two main categories: structured and unstructured pruning. In structured pruning only specific weight patterns can be pruned, e.g. a whole row or column can be removed from the weight matrix, resulting in higher speedup. Unstructured pruning does not have such limitation and is thus more fine-grained, resulting in better accuracy, higher compression rates and better size reduction~\cite{hoefler2021sparsity}.


While pruning techniques have widely been used for natural image analysis~\cite{hoefler2021sparsity}, they have rarely been exploited in the context of medical image processing.
In contrast to former studies that used DNN pruning for image classification~\cite{muckatira2020properties,entezari2020} or image semantic segmentation~\cite{Jeong2021}, in this work, we exploited pruning for nuclei instance segmentation in hematoxylin and eosin (H\&E)-stained histological images. Nuclei instance segmentation plays an essential role in the analysis of histological whole slide images and can be considered as a fundamental step for further analysis~\cite{skinner2017nuclear}. Parameters such as nuclei density or count can be extracted from instance segmentation masks. In the next step this information is used for disease detection, diagnosis and treatment planning~\cite{doi:10.1002/cyto.a.23175}. The most promising approaches for automatic nuclei instance segmentation are based on supervised learning~\cite{monuseg,mahbod2021cryonuseg_org}. Localization-based methods, ternary segmentation approaches and regression-based algorithms are among the most promising models for nuclei instance segmentation~\cite{bancher2021improving,monuseg,10.1007/978-3-030-23937-4_9,graham2019hover}. 

In this study, we utilised a recently developed model for nuclei instance segmentation~\cite{10.1007/978-3-030-23937-4_9}. The model has two main branches, namely a semantic segmentation branch and a deep regression branch. We also modified the model by incorporating pre-trained ResNet-34 models~\cite{He2016} in the encoder section of both segmentation and regression branches. We investigated the impact of DNN magnitude pruning for two strategies, i.e. networks-wide and layer-wise (see Section~\ref{sec:pruning}) on both branches, and final instance segmentation results. By doing so, we explore the pruning impact on three distinct image analysis tasks, namely semantics segmentation, regression and instance segmentation. To the best of our knowledge, the effect of DNN pruning on the nuclei instance segmentation and deep regression performance has not yet been investigated in the related literature. Evaluated on two publicly available nuclei instance segmentation datasets, namely MoNuSeg~\cite{monuseg} and TNBC~\cite{naylor2018segmentation}, our results show that 80\% of the model's weights can be pruned with only up to 2\,\% drop in the nuclei instance  segmentation performance. 

\section{Method}
\label{sec:method}

\begin{figure}[!ht]
	\centering
	\includegraphics[width=.7\linewidth]{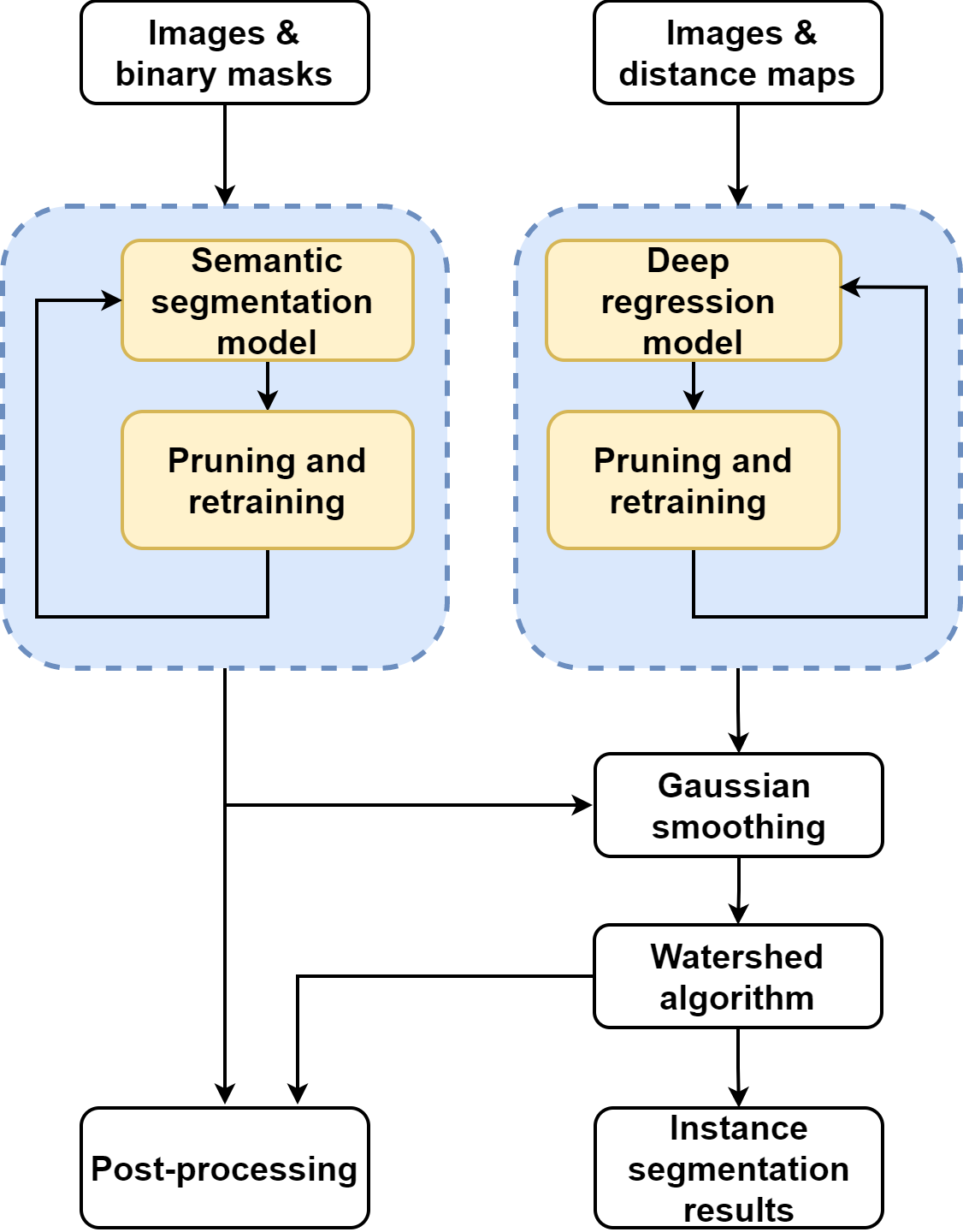}
	\vspace{-10pt}
	\caption{{\bf The generic workflow of the proposed method.} 
		Iterative pruning and training for semantic segmentation and regression models are shown in blue boxes.  
	}
	\label{flowchart}
	\vspace{-10 pt}
\end{figure}

The workflow of a DNN pruning approach for nuclei instance segmentation is depicted in Fig.~\ref{flowchart}. The left branch does semantic segmentation while the right branch performs deep regression. To compress the model, each branch is trained and pruned iteratively. Then, the final compressed networks for both branches are merged together for the final post-processing of the resulting instance segmentation model. 
In the following subsections, we first describe the datasets used, and detailed model in each branch. We then explain the pruning step and the final merging step, which complete the processing pipeline.

\subsection{Datasets}
\label{Datasets}
We investigate the performance of our pruning strategies on two publicly available datasets, MoNuSeg~\cite{monuseg} and TNBC~\cite{naylor2018segmentation}. MoNuSeg dataset consists of 44 H\&E-stained histological images with a fixed image size of 1000$\times$1000 pixels. The image patches were extracted from the TCGA database and nine human organs (breast, kidney, liver, prostate, lung, bladder, colon, stomach, and brain). All images were acquired at $40\times$ magnification, and the entire dataset has more than 28,000 manually segmented nuclei. As described in \cite{monuseg}, the dataset was divided into a training (30 images) and a test set (14 images). Further details about the  MoNuSeg dataset can be found in \cite{monuseg}. TNBC dataset consists of 50 H\&E-stained histological images with a fixed image size of 512$\times$512 pixels. The whole slide images were acquired and scanned at $40\times$ magnification by Curie Institute, Paris, France. The dataset consists of images from one organ (breast). More than 4,000 nuclei were manually segmented in this dataset. Further details are provided in \cite{naylor2018segmentation}. 


\subsection{Segmentation and regression models}
\label{CNN models}
Our instance segmentation model for nuclei segmentation is inspired by~\cite{10.1007/978-3-030-23937-4_9}. As shown in Fig.~\ref{flowchart}, the proposed pruning pipeline processes two models separately, namely a semantic segmentation model and a deep regression model. The main task of the semantic segmentation model to  separate the background from the foreground, while the main task of the deep regression model is to predict the nuclei distance maps. The architectures of the utilised models are the same. Both are encoder-decoder-based models with skip connections between encoder and decoder parts. In contrast to the original architecture presented in \cite{10.1007/978-3-030-23937-4_9}, which uses four convolutional and max-pooling layers in the encoder part, we integrate the ResNet-34 architecture~\cite{He2016} as the encoder for both semantic segmentation and deep regression models as using pre-trained models in the encoder part of the U-Net-alike architectures has shown to improve the performance~\cite{monuseg}.  
Both models have the same number of trainable parameters, which makes around 24.4 million parameters in total.  As the semantic segmentation model dealt with a binary segmentation task, we used the sigmoid activation function in the last layer with the Dice loss function. For the deep regression model, however, we use a linear activation function in the last layer with a mean square error loss function to handle the regression task. Aside from these two differences, the model architectures, the training and the inference procedures are identical in both branches. We use Adam optimizer with a batch size of 2 for model training. The initial learning rate was set to 0.001 with a cosine annealing learning rate scheduler~\cite{loshchilov2016sgdr}. We trained each model for 1000 epochs.  As the DNN models were trained with limited training samples, we made use of a number of augmentation techniques suggested in prior studies~\cite{mahbod2021cryonuseg_org,10.1007/978-3-030-23937-4_9}. We applied resizing (only for the MoNuSeg dataset to match them to 1024$\times$1024 pixels), random horizontal flipping, random scaling and shifting, random Gaussian filtering, and random perspective transformation as the main augmentation techniques.

\subsection{Pruning}
\label{sec:pruning}

Existing literature suggests multiple ways to make use of sparsity during and after model training. \cite{hoefler2021sparsity} provides a comprehensive survey on model sparsification strategies.
Among all these methods, magnitude pruning has gained a lot of attention mainly for two reasons: it features superior performance than many other pruning strategies, and 
it is computationally inexpensive, i.e. there is no need to compute Hessians, etc. In this work, we employ two versions of magnitude pruning, namely network-wide (also known as global) and layer-wise magnitude pruning. Algorithm~\ref{mag_prune} illustrates the pseudo-code for these two pruning techniques. Related work~\cite{renda2020comparing} shows that iterative pruning achieves better performance than one-shot pruning.

\begin{algorithm}[t]
	\label{mag_prune}
	\caption{Network-wide / layer-wise magnitude pruning}
	\begin{algorithmic}[1]
		\Procedure{Iter-Mag-Prune}{$p\_type$, $CR$}       
		
		\State Train neural network until convergence
		\State $i = \log_{2} CR$ \Comment{number of iterations}
		
		\While{$i \geq 1 $}
		\If{$p\_type$ = network-wide}
		\State Sort all weights in ascending order
		\State Prune the smallest 50\% of weights
		
		\ElsIf{$p\_type$ = layer-wise}
		\State Sort parameters in each layer
		\State Prune the smallest 50\% of weights per layer
		\EndIf
		\State Retrain the network until convergence
		\State $i = i - 1$
		\EndWhile  
		\EndProcedure
	\end{algorithmic}
\end{algorithm}

\subsection{Merging and post-processing}
To form the final instance segmentation masks, we merged the results from the semantic segmentation and deep regression branches as shown in Fig.~\ref{flowchart}. The merging scheme is similar to~\cite{10.1007/978-3-030-23937-4_9}. First, we apply a Gaussian smoothing filter on the predicted nuclei distance maps from the deep regression model to prevent false-positive local maxima detection. The estimated average nuclei size from the segmentation model determines the kernel size of the Gaussian filter~\cite{10.1007/978-3-030-23937-4_9}. From the smoothed distance maps, we then extract local maxima and use them as seed points for the marker-controlled Watershed  algorithm. 
We use the predicted binary segmentation masks from the semantic segmentation model to determine the background in the instance segmentation masks. Finally, we apply two straightforward post-processing steps as suggested in former studies~\cite{monuseg}. Using morphological operations, we remove tiny objects from the segmentation masks (area \textless 30 pixels) and fill the holes in the predicted instances.

\subsection{Evaluation Metrics}
To evaluate the final instance segmentation performance, we use Aggregate Jaccard Index (AJI)~\cite{monuseg} and Panoptic Quality (PQ) score~\cite{graham2019hover} as the primary evaluation scores. Moreover, to evaluate the performance of each DNN model (i.e. the semantic segmentation model and the deep regression model), we utilise Dice similarity score and Mean Square Error (MSE) as evaluation indexes. Further details about these scores can be found in~\cite{graham2019hover}.

\section{Results \& Discussion}
\label{sec:res}
\begin{figure}[t]
	\centering
	\includegraphics[width=0.49\columnwidth]{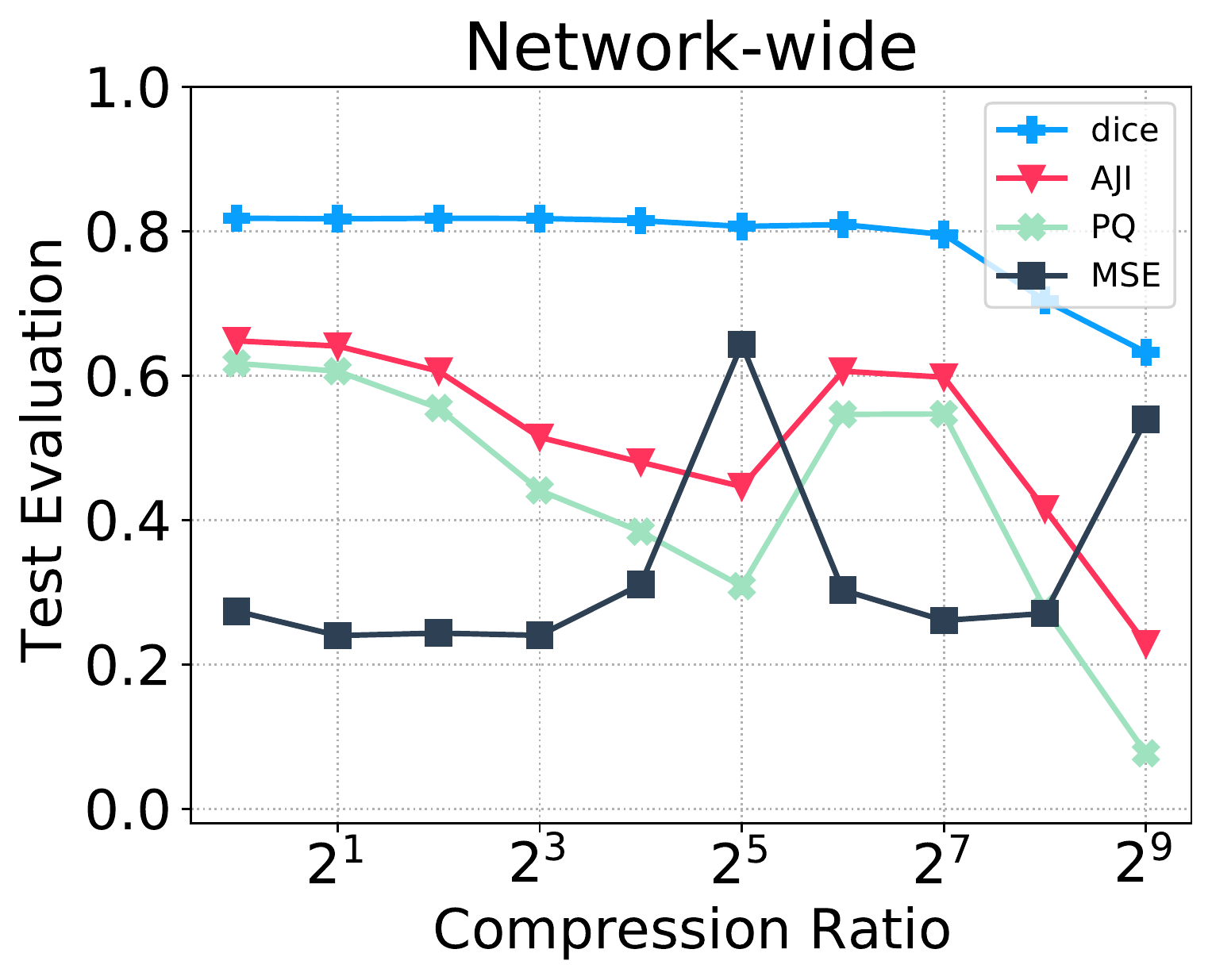}
	\includegraphics[width=0.49\columnwidth]{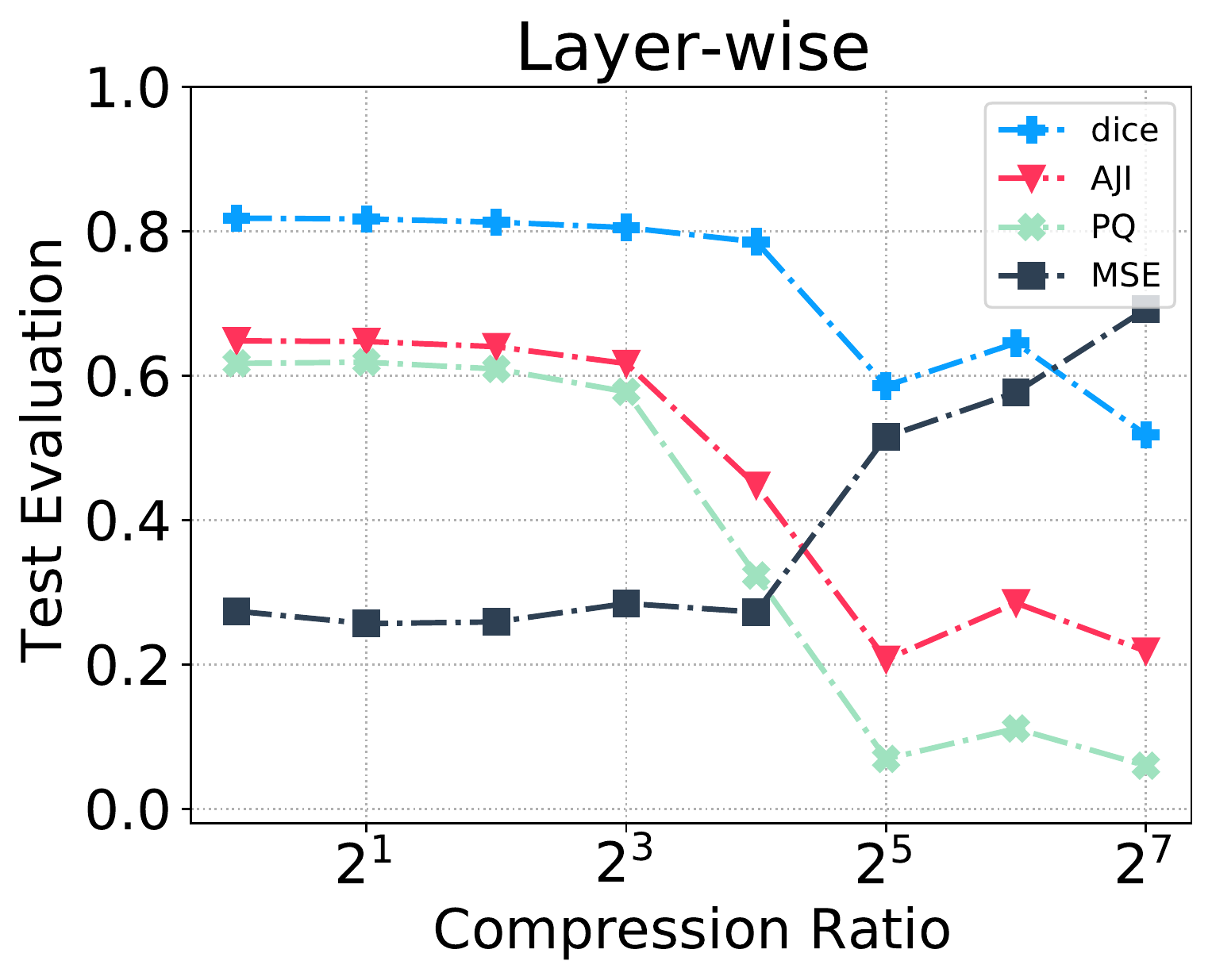}
	\vspace{-10pt}
	\caption{\textbf{Impact of two pruning methods on the semantic segmentation (Dice), regression (MSE), and instance segmentation (AJI, PQ) performances on the MoNuSeg test set.} Nuclei semantic segmentation is extremely robust against both pruning methods.
		Layer-wise pruning shows better performance for smaller compression ratios, while enforcing larger pruning ratios to each layer harms the total performance. 
	}
	\label{monuseg_res}
\end{figure}

\begin{figure}[t]
	\centering
	\includegraphics[width=0.49\columnwidth]{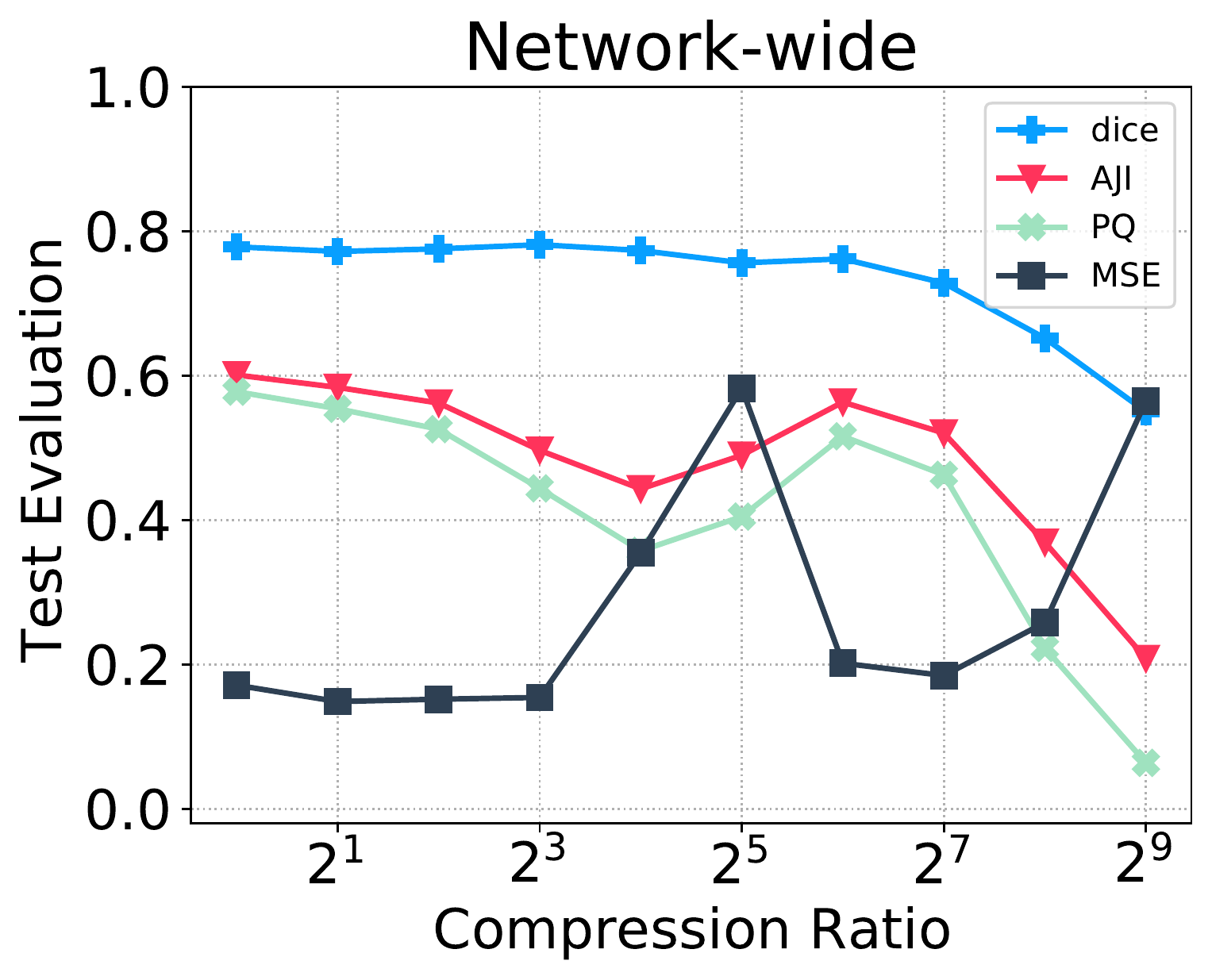}
	\includegraphics[width=0.49\columnwidth]{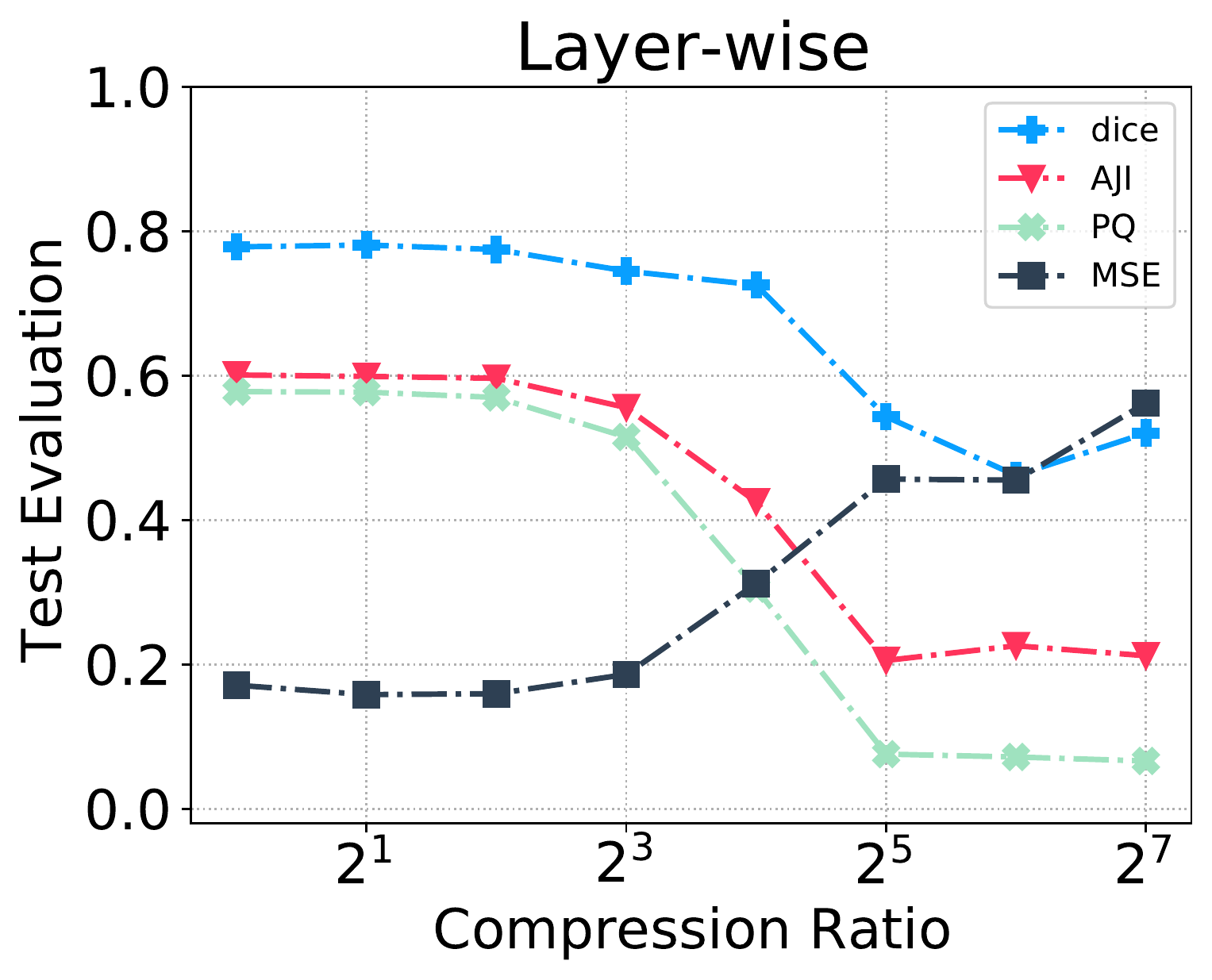}
	\vspace{-10pt}
	\caption{\textbf{Impact of pruning on model performance under distribution shift}, i.e. training and pruning on MoNuSeg and test on TNBC dataset. All evaluation metrics are identical to Fig.~\ref{monuseg_res}. Layer-wise pruning shows less robustness to the distribution shift.
	}
	\label{tnbc_ref}
\end{figure}

In our experiments, we use the training set of the MoNuSeg dataset to train and iteratively prune the model. Fig.~\ref{monuseg_res} and Fig.~\ref{tnbc_ref} present performance results of the pruned models on the
test set of the MoNuSeg dataset and the entire TNBC dataset, respectively.

\begin{table*}[!h]
	\scriptsize
	\setlength\tabcolsep{0pt} 
	\begin{threeparttable}
		
		\caption{{\bf Theoretical Speedup.} 
			Layer-wise pruning yields higher speed-ups, as it prunes more weights in the early layers that have small kernels applied to a large input, requiring more FLOPS. $CR>2^7$ is impossible to achieve with layer-wise pruning due to loss of connectivity inside the model.}
		\label{tab:speedup}
		\begin{tabular*}{\textwidth}{l@{\extracolsep{\fill}}*{9}{c}}
			\toprule
			Pruning Method & \multicolumn{9}{c}{Compression Ratio (CR)}\\
			\cmidrule{2-10}
			& 2 & 4 & 8 & 16 & 32 & 64 & 128 & 256 & 512\\
			\midrule
			Network-wide  & 1.49 & 2.13 & 3.07 & 4.46 & 6.59 & 9.90 & 15.01 & 23.80 & 45.02\\
			Layer-wise  & 1.99 & 3.99 & 7.97 & 15.89 & 31.57 & 62.27 & 121.18 & - & -\\
			\bottomrule 
		\end{tabular*}  
		
	\end{threeparttable}
\end{table*}

\begin{figure*}[t]
	\centering
	\includegraphics[width=0.95\textwidth]{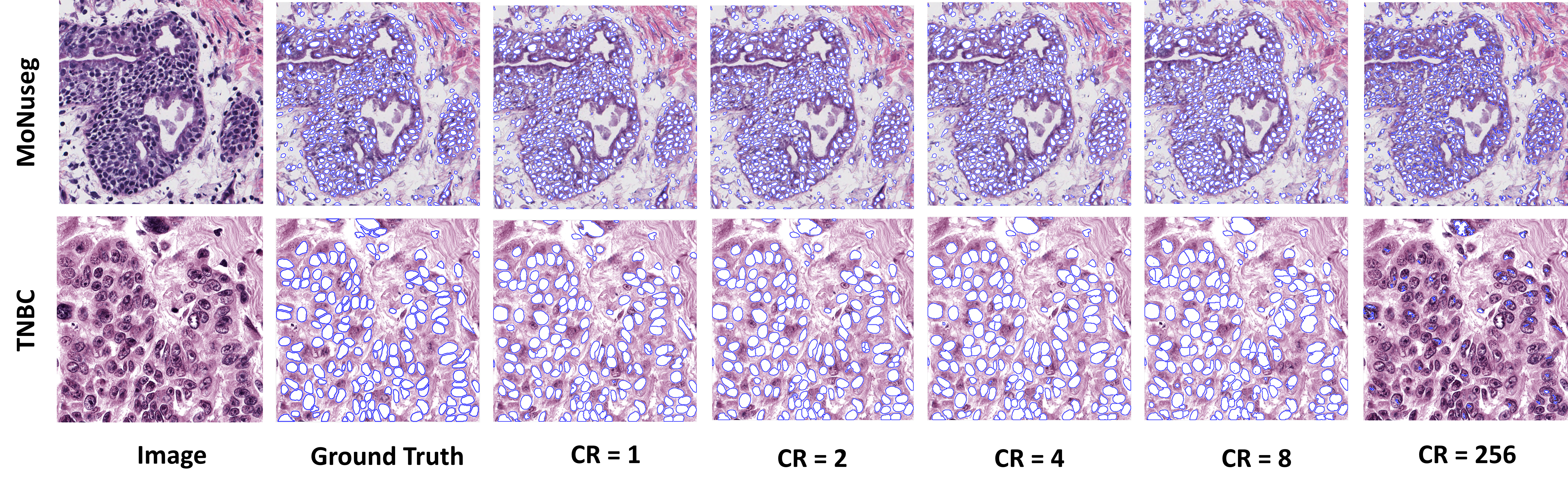}
	\vspace{-15pt}
	\caption{\textbf{Visualizing the layer-wise pruning impact on the final instance segmentation performance} for two example images from MoNuSeg (top) and TNBC dataset (bottom). The first column shows the raw input images, the second column shows the ground truth instance segmentation masks, and the rest shows the predicted masks by the model with different compression ratios (CRs).
	}
	\label{examples}
	\vspace{-5pt}
\end{figure*}

Fig.~\ref{monuseg_res} shows that nuclei semantic segmentation network is extremely robust against pruning, where removing $0.992\%$ of the parameters in network-wide ($CR=2^7$) and $0.937\%$ ($CR=2^4$) in layer-wise fashion is possible with only 2\% reduction in Dice. It also shows that layer-wise pruning is a better choice (for all measures and tasks) for smaller compression ratios, while enforcing extreme pruning ratios ($CR>2^5$) to each layer harms the total performance. That is due to losing small yet important kernels in the early layers. Extreme pruning ratios ($CR>2^8$) are also not possible to achieve with layer-wise pruning because this leads to removing the whole layer.

Another interesting observation in Fig.~\ref{monuseg_res} (network-wide) is that pruning deep regression model first increase MSE and then decrease MSE (and ultimately results in high MSE values due to extreme compression ratios). This observation is aligned to previous works~\cite{renda2020comparing}, where the authors observe the performance improvement when pruning. We conjecture that iterative pruning and training helps to find better global optima in the loss landscape. Such a performance improvement results in extreme compression ratios ($CR=2^7$) for network-wide instance segmentation pruning while only losing 2\,\% in AJI and PQ.

Fig.~\ref{tnbc_ref} evaluates model performance under distribution shift, i.e. training and pruning on MoNuSeg and test on the TNBC dataset. All evaluation metrics are close to Fig.~\ref{monuseg_res}, showing that our pruning schemes are robust to potential natural distribution shifts presented by a different dataset. Comparing network-wide and layer-wise pruning shows that the latter presents less robustness to the distribution shift. Similar evaluation patterns also can be seen in the TNBC dataset, i.e. for smaller CRs, layer-wise is a better choice for instance segmentation. We also observe the performance decrease and then increase for network-wide pruning, as better solutions are found using iterative pruning and training.

Table.~\ref{tab:speedup} shows the theoretical speedup that can potentially be achieved by proper hardware supporting model sparsity for both pruning techniques along different CRs. As expected, layer-wise pruning gains much higher speed-ups, as it stronger prunes early  layers. 

Fig.~\ref{examples} provides qualitative evaluation of the impact of pruning on the final instance segmentation for different CRs. DNN pruning with small CRs (CR $\approx$ $2^3$ and smaller) has not drastically changed the predicted instance segmentation masks. However, the instance segmentation performance has been significantly degraded for very large CR ($CR=2^8$). 

This work can be extended in a number of ways that we are planning to address in future studies. We plan to investigate the effect of architecture on instance segmentation by using different pre-trained models. 
Another interesting direction is to investigate the observed distribution shift for different target datasets. It is also interesting to see whether fine-tuning on the target distribution helps the out-of-distribution generalization. Another potential research question, inspired by \cite{hooker2019compressed} is to investigate which instances (classes) are more prone to neural network compression. Finally, it would be interesting to explore structured pruning techniques, especially those which allow practical speedups when using special hardware such as NVIDIA Ampere~\cite{zhou2021learning}. 

\section{Conclusion}
\label{conclusion}
In this work, we apply two magnitude-based pruning techniques, namely network-wide and layer-wise pruning for nuclei instance segmentation in H\&E-stained histological images. Our results suggest that with layer-wise pruning and a certain CR level, the weights from the semantic segmentation and deep regression models can be pruned with less than 2\% drop in the evaluation indexes. We observe that nuclei semantic segmentation is highly robust against pruning i.e. Dice score is barely reduced even in extreme compression ratios. Our results also shows that both pruning methods show high robustness again distribution shift. Further research is needed to explore other pruning techniques and investigate their impact on the nuclei instance segmentation performance for in-distribution and out-of-distribution regimes.  


\subsubsection{Acknowledgements} This  project was  supported by  the  Austrian Research Promotion Agency (FFG), No. 872636.
This study was conducted retrospectively using human subject data made available through open access. Ethical approval was not required as confirmed by the license attached with the open access data.

%
%
%
\bibliographystyle{splncs04}
\bibliography{miccai.bib}

\end{document}